\documentstyle[letter,epsf]{ptptex}

\newcommand{\gsim}{\raisebox{-0.55ex}{$\stackrel{\displaystyle >}{\sim}$}}
\newcommand{\diff}[2]{\frac{\partial #1}{\partial #2}}

\notypesetlogo  

\title{
$\eta$-Scaling of $dN_{\rm ch}/d\eta$ at $\sqrt{s_{NN}} = 200\ {\rm GeV}$ by the PHOBOS Collaboration and the Ornstein-Uhlenbeck stochastic process
}

\author{
Minoru Biyajima$^{1,2}$ and Takuya Mizoguchi$^3$
}

\inst{
$^1$Department of Physics, Faculty of Science, Shinshu University, Matsumoto 390-8621, Japan\\

$^2$Niels Bohr Institute, DK-2100, Copenhagen, Denmark\\

$^3$Toba National College of Maritime Technology, Toba 517-8501, Japan}

\recdate{
\today
}

\abst{
Using the latest data concerning $dN_{\rm ch}/d\eta$ at $\sqrt{s_{NN}} = 200$ GeV, we have analyzed them by means of stochastic theory named the Ornstein-Uhlenbeck process with two sources. Moreover, we display that $z_r = \eta/\sqrt{\langle \eta^2\rangle} = \eta/\eta_{\rm rms}$ scaling with centrality cuts are described by two Gaussian distributions.
}

\begin{document}
\maketitle
1. {\it Introduction}\quad In previous papers,\cite{Biyajima:2002at} we have shown that there is an $\eta$-scaling in the charged multiplicity distributions $dN_{\rm ch}/d\eta$ or $(N_{\rm ch})^{-1}dN_{\rm ch}/d\eta = dn/d\eta$ at $\sqrt{s_{NN}} = 130$ GeV by PHOBOS Collaboration.\cite{Back:2001bq} Those distributions have been explained by a stochastic approach named the Ornstein-Uhlenbeck (OU) process with the evolution parameter $t$, the frictional coefficient $\gamma$ and the variance $\sigma^2$ : 
\begin{eqnarray}
  \diff{P(y,\, t)}t = \gamma \left[\diff{}yy + \frac 12\frac{\sigma^2}{\gamma}\diff{^2}{y^2}\right] P(y,\, t)\:.
\label{eq1}
\end{eqnarray}
Introducing two sources at $\pm y_{\rm max} = \ln (\sqrt{s_{NN}}/m_N)$ at $t=0$ and $P(y,\, 0) = 0.5[\delta (y + y_{\rm max})+\delta (y - y_{\rm max})]$, we obtain the following distribution function for $dn/d\eta$ (assuming $y \approx \eta$) using the probability density $P(y,\;t)$\cite{Biyajima:2002at}
\begin{eqnarray}
\frac{dn}{d\eta} = \frac 1{\sqrt{8\pi V^2(t)}}\left\{
\exp\left[-\frac{(\eta+\eta_{\rm max}e^{-\gamma t})^2}{2V^2(t)}\right]
+ \exp\left[-\frac{(\eta-\eta_{\rm max}e^{-\gamma t})^2}{2V^2(t)}\right]\right\},\quad
\label{eq2}
\end{eqnarray}
where $V^2(t) = (\sigma^2/2\gamma)p$ with $p = 1-e^{-2\gamma t}$.

On the other hand, in Ref. \citen{Nouicer:2002ks}, PHOBOS Collaboration has reported new analyses at 200 GeV. Thus, we are interested in comparisons of our results in Ref. \citen{Biyajima:2002at} with theoretical analyses of data in Ref.~\citen{Nouicer:2002ks}, in particular Eq.~(\ref{eq2}) and its $z_r = \eta/\eta_{\rm rms}$ ($\eta_{\rm rms}=\sqrt{\langle\eta^2\rangle}$) scaling function with $z_{\rm max} = \eta_{\rm max}/\eta_{\rm rms}$ and $V_r^2(t) = V^2(t)/\eta_{\rm rms}^2$.
\begin{eqnarray}
\frac{dn}{dz_r} = \frac 1{\sqrt{8\pi V_r^2(t)}}\left\{\exp\left[-\frac{(z_r+z_{\rm max}e^{-\gamma t})^2}{2V_r^2(t)}\right]+ \exp\left[-\frac{(z_r-z_{\rm max}e^{-\gamma t})^2}{2V_r^2(t)}\right]\, \right\}.\quad
\label{eq3}
\end{eqnarray}
First we have to examine $\eta$ scaling of $dn/d\eta$ in a semi-phenomenological point of view. After analyses in terms of Eqs.~(\ref{eq2}) and (\ref{eq3}), we examine the meaning of the evolution parameter $t$ with $\gamma$, assigning a physical dimension [second] to it. Finally concluding remarks are given.

\begin{figure}[htb]
  \epsfxsize= 10 cm   
  \centerline{\epsfbox{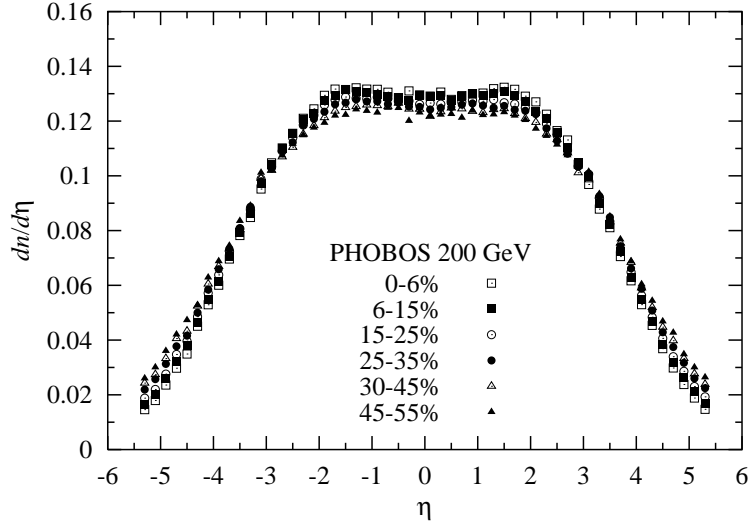}}
  \caption{$dn/d\eta$ with different centrality cuts.\cite{Nouicer:2002ks}}
\label{fig1}
\end{figure}

2. {\it Semi-phenomenological analyses of data}\quad In Fig.~\ref{fig1}, we show distributions of $dn/d\eta$. As is seen in Fig.~\ref{fig1}, the intercepts of $dn/d\eta|_{\eta = 0}$ with different centrality cuts are located in the following narrow interval
\begin{eqnarray}
\left. \frac{dn}{d\eta}\right|_{\eta = 0} = c = 0.123-0.130\:.
\label{eq4}
\end{eqnarray}
We can observe that the $\eta$-scaling approximately holds at $\sqrt{s_{NN}} = 200$ GeV.

\begin{figure}[htb]
\begin{tabular}{cc}
\begin{minipage}{.47\hsize}
  \epsfxsize= 6.6 cm   
  \centerline{\epsfbox{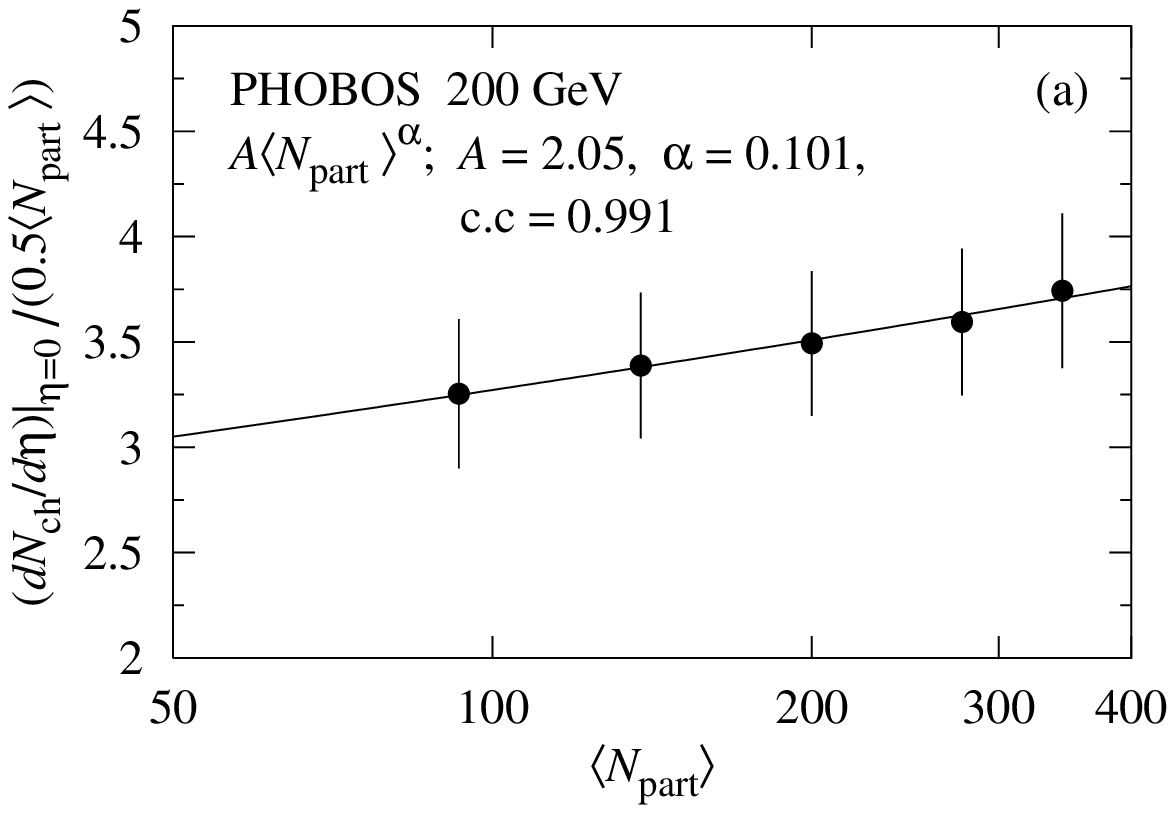}}
\end{minipage}
&
\begin{minipage}{.47\hsize}
  \epsfxsize= 6.6 cm   
  \centerline{\epsfbox{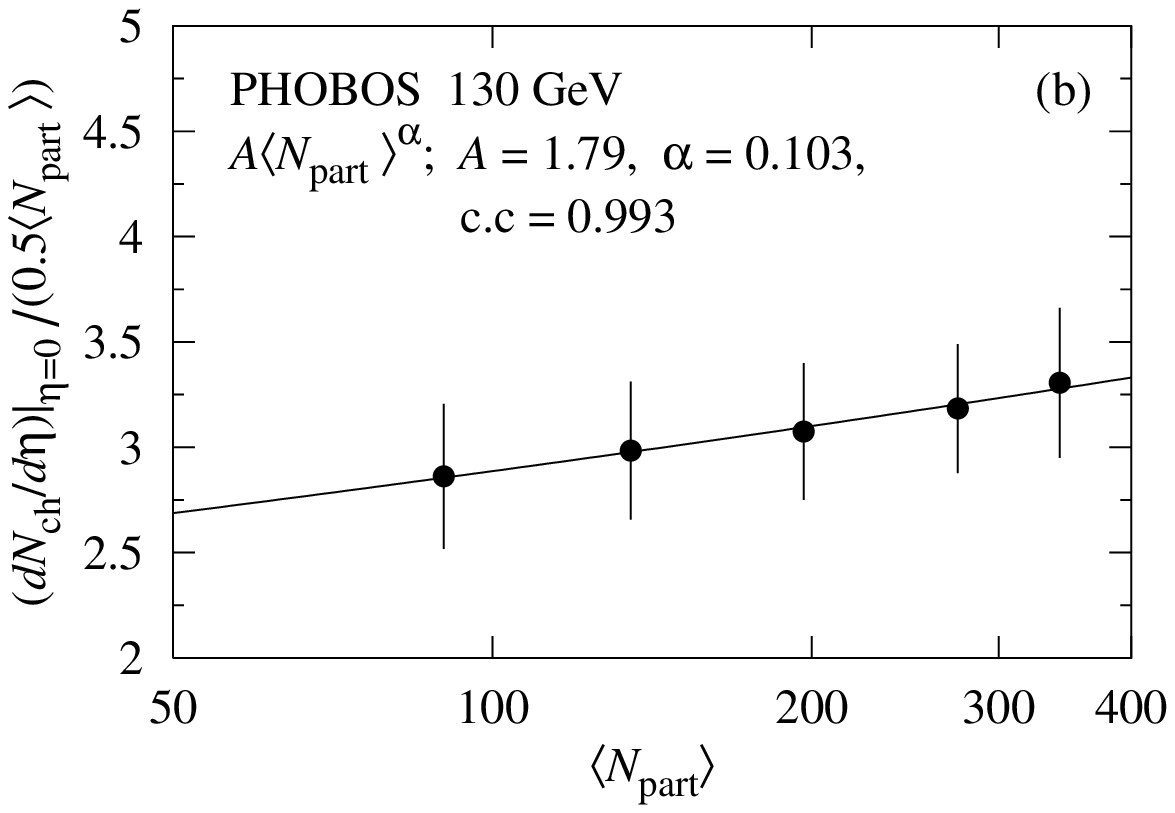}}
\end{minipage}
\end{tabular}
  \caption{Determination of the parameters $A$ and $\alpha$. The method of linear regression is used. The correlation coefficient (c.c.) is 0.991 (200 GeV). From data at 130 GeV we have $A=1.79$, $\alpha = 0.103$ and c.c. $=$ 0.993.}
\label{fig2}
\end{figure}

Next, we should examine a power-like law in $(0.5\langle N_{\rm part}\rangle)^{-1}(dN_{\rm ch}/d\eta)|_{\eta = 0}$ ($\langle N_{\rm part}\rangle$ being number of participants) which is proposed by WA98 Collaborations\cite{Aggarwal:2000bc} as
\begin{eqnarray}
\left. \frac 1{0.5\langle N_{\rm part}\rangle}\frac{dN_{\rm ch}}{d\eta}\right|_{\eta = 0} = A\langle N_{\rm part}\rangle^{\alpha}\:.
\label{eq5}
\end{eqnarray}
As seen in Fig.~\ref{fig2}, it can be stressed that the power-like law holds fairly well\footnote{
According to Ref.~\citen{Aggarwal:2000bc}, this law suggests the following picture: The incoming particles lose their memory and every participants contributes a similar amount of energy to particle production.
}. Using estimated parameters $A$ and $\alpha$, we can express $c$ as
\begin{eqnarray}
c = \frac{0.5\langle N_{\rm part}\rangle}{N_{\rm ch}}\times (2.05)\langle N_{\rm part}\rangle^{0.101}\:.
\label{eq6}
\end{eqnarray}
They are shown in Table \ref{table1}. The intercepts show weak decreasing behavior as the centrality cut increases.
%
%
%
%
%
%
\begin{table}[htb]
\begin{center}
\caption{Empirical examination of Eq.~(\ref{eq6}) ($\sqrt{s_{NN}} = 200$ GeV).}
\label{table1}
\begin{tabular}{cccccc} \hline\hline
centrality (\%) & 35--45 & 25--35 & 15--25 & 6--15 & 0--6\\ \hline
$\langle N_{\rm part}\rangle$ & 93$\pm$5 & 138$\pm$6 & 200$\pm$7.5 & 277$\pm$8.5 & 344.5$\pm$11\\
$N_{ch}^{\rm (Ex)}$ & 1230$\pm$60 & 1870$\pm$90 & 2750$\pm$140 & 3860$\pm$190 & 4960$\pm$250\\
$c$ (Ex) & 0.123$\pm$0.012 & 0.124$\pm$0.012 & 0.127$\pm$0.012 & 0.129$\pm$0.012 & 0.130$\pm$0.012 \\
$c$ (Eq.~(\ref{eq6})) & 0.122$\pm$0.009 & 0.124$\pm$0.008 & 0.127$\pm$0.008 & 0.130$\pm$0.008 & 0.128$\pm$0.008\\ \hline
\end{tabular}
\end{center}
\end{table}

3. {\it Analyses of data by Eqs.~(\ref{eq2}) and (\ref{eq3})}

\begin{figure}[htb]
  \epsfxsize= 13 cm   
  \centerline{\epsfbox{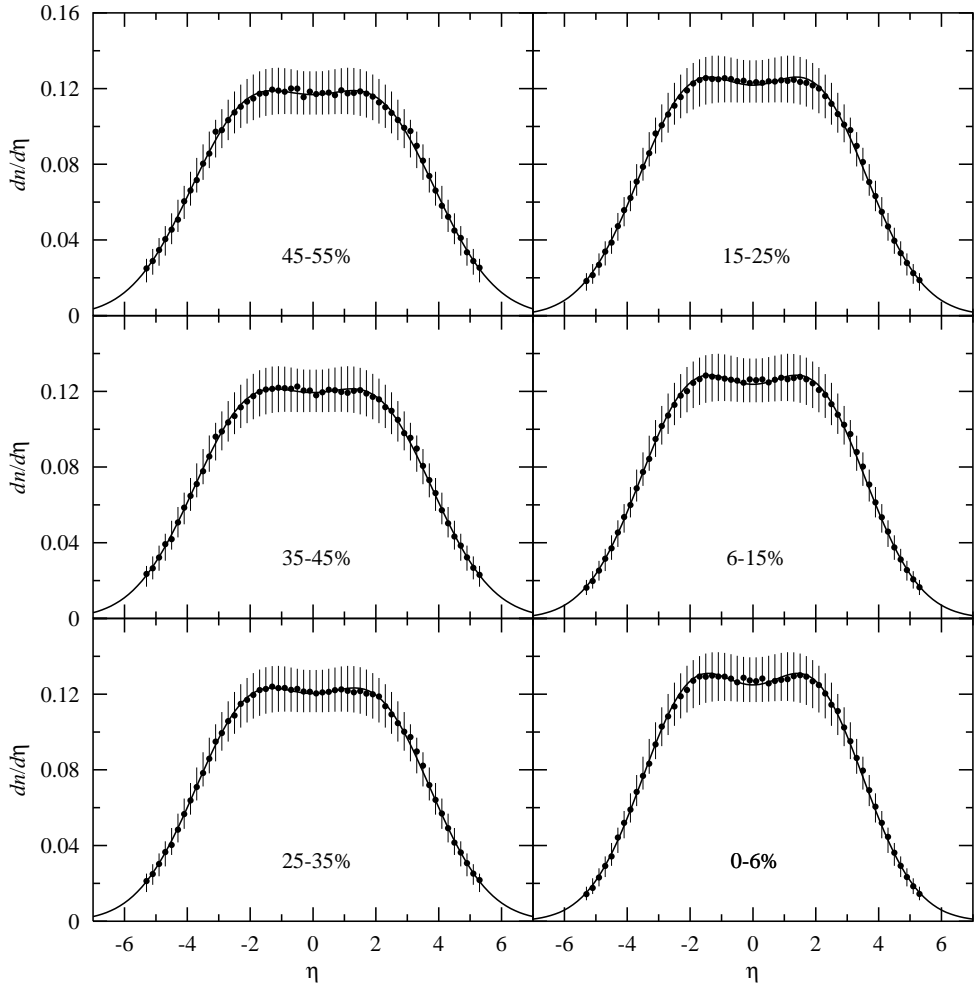}}
  \caption{Analyses of $dn/d\eta$ with centrality cuts using Eq.~(\ref{eq2}). (See Table \ref{table2}.)}
\label{fig3}
\end{figure}

3.1 {\it Explanation of $dn/d\eta$ by Eq.~(\ref{eq2})}\quad This time $\pm \eta_{\rm max} = \pm 5.4$ are taken. The estimated parameters $V^2(t)$, $p$, $c({\rm Th}) = (1/\sqrt{2\pi V^2(t)})\cdot\exp\left[-(\eta_{\rm max}e^{-\gamma t})^2/2V^2(t)\right]$ and $\chi^2$ are shown in Table~\ref{table2}. The results are shown in Fig.~\ref{fig3}. To describe the dip structures, the finite evolution time is necessary in our approach.
%
%
\begin{table}[htb]
\begin{center}
\caption{Parameter values obtained in our analyses using Eq.~(\ref{eq2}). The evolution of $P(y,\, y_{\rm max},\, t)$ in Eq.~(\ref{eq2}), i.e., $dn/d\eta$ ($y\approx \eta$), is stopped at minimum values of $\chi^2$. Here, $\delta p =$ 0.004--0.006 and $\delta c_t =$ 0.005--0.006. (n.d.f. means the number of degree of freedom.) Values of $\eta_{\rm rms}$ are also shown. Notice that $c^*({\rm Th}) = c({\rm Th})\cdot N_{ch}^{\rm (Th)}/N_{ch}^{\rm (Ex)}$. This ratio is needed for correction for $|\eta|>5.4$.}
\label{table2}
\begin{tabular}{ccccccc} \hline\hline
centrality (\%) & 45--55 & 35--45 & 25--35 & 15--25 & 6--15 & 0--6 \\ \hline
$p$ & 0.855$\pm \delta p$ & 0.861$\pm \delta p$ & 0.864$\pm \delta p$ 
    & 0.868$\pm \delta p$ & 0.873$\pm \delta p$ & 0.876$\pm \delta p$\\
$V^2(t)$ & 3.62$\pm$0.26 & 3.49$\pm$0.23 & 3.31$\pm$0.20 
         & 3.09$\pm$0.17 & 2.95$\pm$0.15 & 2.79$\pm$0.13\\
$N_{ch}^{\rm (Th)}$ & 780$\pm$13 & 1270$\pm$21 & 1930$\pm$30 
                    & 2821$\pm$43 & 3951$\pm$60 & 5050$\pm$77\\
$c^*$(Th) 
& 0.121$\pm \delta c_t$ & 0.123$\pm \delta c_t$ & 0.124$\pm \delta c_t$ 
& 0.125$\pm\delta c_t$ & 0.127$\pm \delta c_t$ & 0.127$\pm \delta c_t$\\
$\chi^2/{\rm n.d.f.}$ & 1.07/51 & 0.91/51 & 0.88/51 
                      & 1.18/51 & 1.06/51 & 1.46/51\\
$\eta_{\rm rms} = \sqrt{\langle \eta^2\rangle}$ & 2.57 & 2.54 & 2.51 & 2.48 & 2.45 & 2.42\\ \hline
\end{tabular}
\end{center}
\end{table}
%
%

\begin{figure}[htb]
  \epsfxsize= 13 cm   
  \centerline{\epsfbox{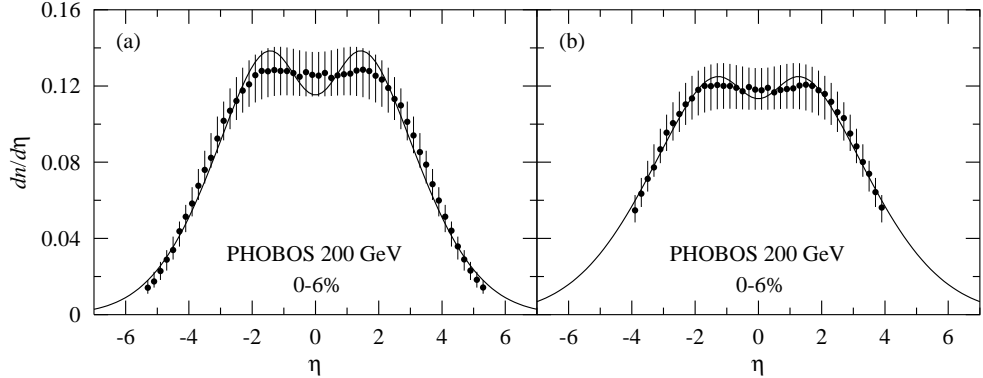}}
  \caption{Analyses of $dn/d\eta$ by means of single Gaussian and Eq.~(\ref{eq7}). (a) Data in full-$\eta$ variable are described by $V^2(t) = 5.27\pm 0.26$ and $m/p_t = 1.13\pm 0.10$. The best $\chi^2 = 20.0/51$. (b) Data in $|\eta|<4.0$ are used, i.e., $4.0<|\eta|<5.4$ are neglected. $V^2(t) = 7.41\pm 0.85$ and $m/p_t = 0.82\pm 0.13$ are used. The best $\chi^2 = 4.0/37$. Introduction of renormalization is necessary, due to the Jacobian}
\label{fig4}
\end{figure}

3.2 {\it Comparison with other approaches}\quad First we consider a problem between the role of Jacobian and dip structure at $\eta \approx 0$. The authors of Refs.~\citen{Kharzeev:2001gp} and \citen{Eskola:2002qz} have explained $dN_{\rm ch}/d\eta$ by means of the Jacobian between the rapidity variable ($y$) and the pseudorapidity ($\eta$): The following relation is well known
\begin{eqnarray}
\frac{dn}{d\eta} = \frac pE \frac{dn}{dy} = \frac{\cosh \eta}{\sqrt{1+ m^2/p_t^2+\sinh^2 \eta}} \frac{dn}{dy}\,,
\label{eq7}
\end{eqnarray}
where $dn/dy = (1/\sqrt{2\pi V^2(t)})\exp\left[-y^2/2V^2(t)\right]$. It is worthwhile to examine whether or not $dn/d\eta$ at $\sqrt{s_{NN}} = 200$ GeV can be explained by Eq.~(\ref{eq7}). As is seen in Fig.~\ref{fig4}, for $dn/d\eta$ in the full phase space ($|\eta|<5.4$), it is difficult to explain the $\eta$ distribution. On the other hand, if we restrict the central region ($|\eta|<4)$, i.e., neglecting the data in $4.0<|\eta|<5.4$, we have better description. These results are actually utilized in Refs.~\citen{Kharzeev:2001gp} and \citen{Eskola:2002qz}. In other words, this fact suggests us that we have to consider other approaches to explain the dip structure in central region as well as the behavior in the fragmentation region. In our case it is the stochastic theory named the OU process with two sources at $\pm y_{\rm max}$ and at $t=0$. 
%
%

\begin{figure}[htb]
\begin{tabular}{cc}
\begin{minipage}{.47\hsize}
  \epsfxsize= 6.6 cm   
  \centerline{\epsfbox{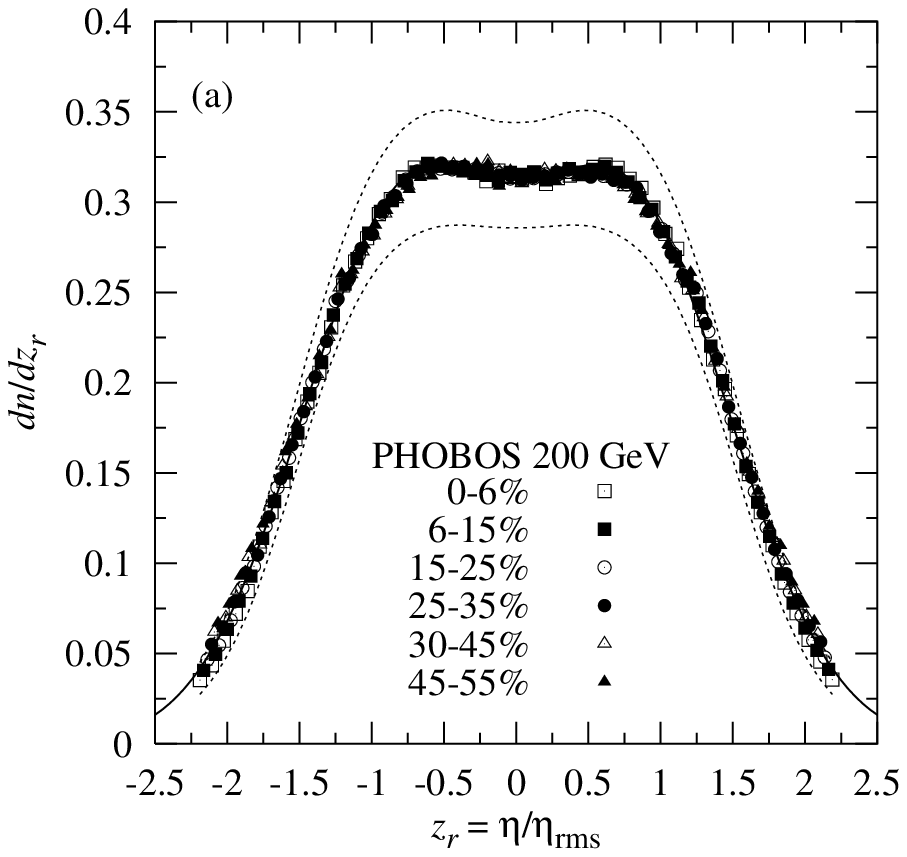}}
\end{minipage}
&
\begin{minipage}{.47\hsize}
  \epsfxsize= 6.6 cm   
  \centerline{\epsfbox{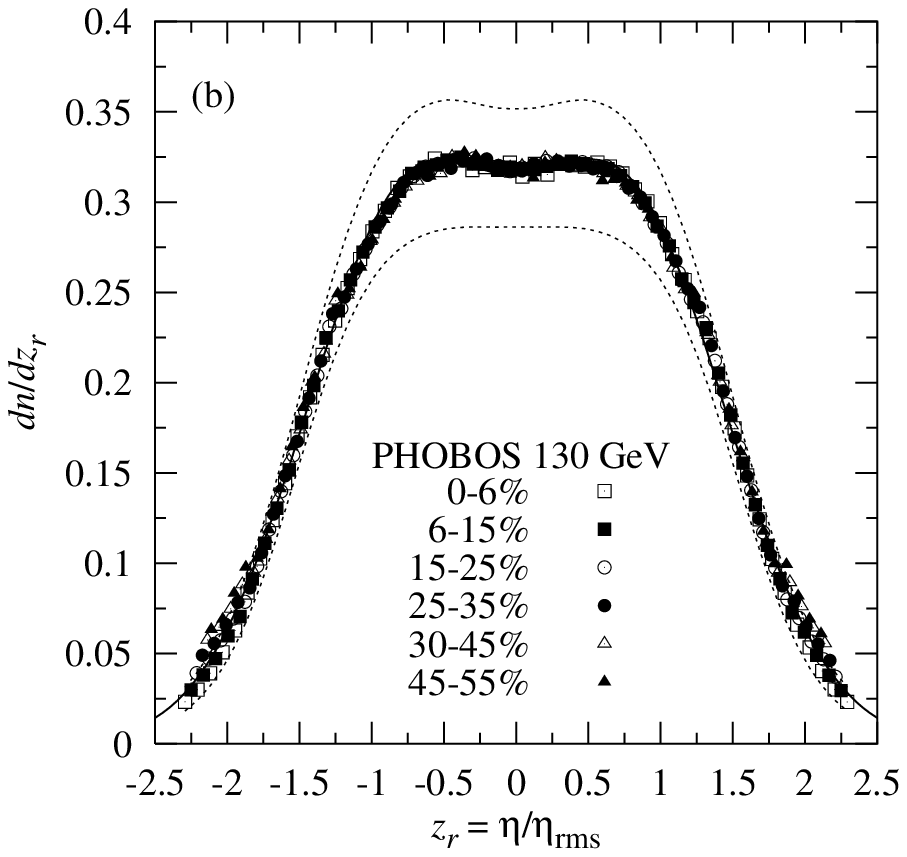}}
\end{minipage}
\end{tabular}
  \caption{Normalized distribution of $dn/dz_r$ with $z_r=\eta/\eta_{\rm rms}$ scaling and estimated parameters using Eq.~(\ref{eq3}). (a) $\sqrt{s_{NN}} = 200$ GeV, $p = 0.867\pm 0.002$, $V_r^2(t) = 0.506\pm 0.011$, $\chi^2/{\rm n.d.f.} = 14.6/321$. (b) $\sqrt{s_{NN}} = 130$ GeV, $p = 0.854\pm 0.002$, $V_r^2(t) = 0.494\pm 0.010$, $\chi^2/{\rm n.d.f.} = 25.5/321$. Dashed lines are magnitudes of error-bars. Notice that $\langle z_r^2\rangle = z_{\rm max}(1-p)+V_r^2 \ne 1.0$, due to the sum of two Gaussian distributions.}
\label{fig5}
\end{figure}

3.3 {\it $z_r$ scaling in $dn/d\eta$ distributions}\quad The values of $\eta_{\rm rms} = \sqrt{\langle \eta^2\rangle}$ are calculated in Table~\ref{table2}. The $z_r = \eta/\eta_{\rm rms}$ distributions are shown in Fig.~\ref{fig5}(a). As seen there, $z_r$ scaling in $dn/d\eta$ distributions with different centrality cuts holds at $\sqrt{s_{NN}} = 200$ GeV. To compare $z_r$ scaling at 200 GeV with one at 130 GeV,\cite{Biyajima:2002at} we show the latter in Fig.~\ref{fig5}(b). It is difficult to distinguish them. This coincidence means that there is no change in $dn/dz_r$ as colliding energy increases, except for the region of $|z_r|\gsim 2.2$.
%
%

4. {\it Interpretation of the evolution parameter $t$ with $\gamma$}\quad In our present treatment the evolution parameter $t$ and the frictional coefficient $\gamma$ are dimensionless. When we assign the meaning of second [s] to $t$, the frictional coefficient $\gamma$ has the dimension of [sec$^{-1} = (1/3)\times 10^{-23}$ fm$^{-1}$]. The magnitude of the interaction region in Au-Au collision is assumed to be 10 fm. See, for example, Ref.~\citen{Morita:2002av}. $t$ is estimated as 
\begin{eqnarray}
t \approx 10\;{\rm fm}/c \approx 3.3\times 10^{-23}\; {\rm sec}\:.
\label{eq8}
\end{eqnarray}
The frictional coefficient and the variance are obtained in Table \ref{table3}. They are comparable with values [$\tau_{\tiny Y}^{-1} = 0.1-0.08$ fm$^{-1}$] of Ref.~\citen{Wolschin:1999jy}, which have been obtained from the proton spectra at SPS collider.

%
%
\begin{table}[htb]
\begin{center}
\caption{Values of $\gamma$ and $\sigma^2$ at $\sqrt{s_{NN}} = 200$ GeV provided that $t\approx 3.3\times 10^{-23}$ sec.}
\label{table3}
\begin{tabular}{cccccccc} \hline\hline
centrality (\%) & 45--55 & 35--45 & 25--35 & 15--25 & 6--15 & 0--6 & average\\ \hline
$\gamma$ [fm$^{-1}$] & 0.096 & 0.099 & 0.100 & 0.101 & 0.103 & 0.104 & 0.101\\
$\sigma^2$ [fm$^{-1}$] & 0.817 & 0.800 & 0.763 & 0.720 & 0.696 & 0.666 & 0.744\\$\sigma^2/\gamma$ & 8.51 & 8.08 & 7.63 & 7.13 & 6.76 & 6.40 & 7.42\\ \hline
\end{tabular}
\end{center}
\end{table}

5. {\it Concluding remarks}\quad {\it c1)} We have analyzed $dn/d\eta$ distribution by Eqs.~(\ref{eq2}) and (\ref{eq6}). The intercepts $c$'s do not show remarkable energy dependence\footnote{
We have observed that the intercepts $c$ are decreasing as the colliding energy increases: $c$ (130 GeV) $=$ 0.125--0.138 $\to$ $c$ (200 GeV) $=$ 0.123--0.130. From the semi-phenomenological analyses, i.e., Eq.~(\ref{eq6}), the reason is attributed to the changed ratio $A^{(200\; {\rm GeV})}/A^{(130\; {\rm GeV})} = 1.15$ and decreased ratio $N_{\rm ch}$(130 GeV)/$N_{\rm ch}$(200 GeV) $\approx$ 0.82. $1.15\times 0.82 \approx 0.94$. On the other hand, in the view of the stochastic approach, i.e., Eq.~(\ref{eq2}), the reason is mainly attributed to the $\eta_{\rm max} = \ln (\sqrt{s_{NN}}/m_N)$ as 
$$
\frac{c\ (200\ {\rm GeV})}{c\ (130\ {\rm GeV})} = \frac{V(t)^{(200)}}{V(t)^{(130)}}\exp\left[-\frac{\eta_{\rm max}^{2(200)}}{2V^2(t)^{(200)}} + \frac{\eta_{\rm max}^{2(130)}}{2V^2(t)^{(130)}}\right] \approx 0.94\:,
$$
where the suffixes mean colliding energies.}.

\noindent
{\it c2)} We have shown that the dip structures in $dn/d\eta$ are hardly explained only by the Jacobian factor from $dn/dy$ in Fig.~\ref{fig4}. What the authors of Refs.~\citen{Kharzeev:2001gp} and \citen{Eskola:2002qz} have stressed is the explanation on the dip structures of $dn/d\eta$ in the central region ($|\eta|<4.0$) by the Jacobian factor. This situation should be changed in analyses of $dn/d\eta$ in the whole pseudo-rapidity region. (See Fig.~\ref{fig4}.)

\noindent
{\it c3)} From the evolution parameter $t$ with $\gamma$ using the assumption for the size of interaction ($R({\rm Au}+{\rm Au}) \approx 10$ fm), we have estimated the frictional coefficient which is almost consistent with values at SPS energies in Ref.~\citen{Wolschin:1999jy}

\noindent
{\it c4)} From $dn/dz_r$'s at $\sqrt{s_{NN}} =$ 200 GeV and 130 GeV, we have shown that both distributions are coincided with each other. If there are no labels (200 GeV and 130 GeV) in Fig.~\ref{fig5}, we cannot distinguish them. This coincidence means that there is no particular change in $dn/d\eta$ between $\sqrt{s_{NN}} =$ 200 GeV and 130 GeV.

%
%
\noindent
\\
{\it Acknowledgements}\quad The authors would like to thank Dr. R.~Nouicer for his kind communication concerning the data. Various conversations with A.~Bialas, H.~Boggild, J.~P.~Bondorf, T.~Csorgo, T.~Hirano, R.~Hwa, H.~Ito, S.~Muroya, K.~Tuominen and G.~Wolschin are very useful for the present study. One of authors~(M.B.) would like to acknowledge The Scandinavia-Japan Sasakawa Foundation for financial support to his stay at Niels Bohr Institute.

%

%
\end{document}